\documentclass[oneside,11pt]{article}%
\usepackage{amsthm}
\usepackage{amsfonts}
\usepackage{amsmath}
\usepackage{amssymb}
\usepackage[colorlinks=true,linkcolor=RawSienna,citecolor=RawSienna,urlcolor=RawSienna,bookmarksopen=true,pdfstartview=FitB]{hyperref}
\usepackage{natbib}
\usepackage{float}
\usepackage{graphicx}
\usepackage[dvipsnames]{xcolor}
\usepackage{footmisc}
\usepackage[left=1.1in,right=1.1in,top=1.1in,bottom=1.1in]{geometry}
\usepackage[onehalfspacing]{setspace}
\usepackage[ruled,vlined]{algorithm2e}
\usepackage{hyphenat}
\newtheorem{theorem}{Theorem}[]

\allowdisplaybreaks
\begin{document}

\title{\vspace{-1.5cm}A weighted FDR procedure under discrete and heterogeneous null distributions}
\author{Xiongzhi Chen\thanks{Corresponding author: Department of Mathematics and
Statistics, Washington State University, Pullman, WA 99164, USA; Email:
\texttt{xiongzhi.chen@wsu.edu}.}, \ Rebecca W. Doerge\thanks{Office of the
Dean, Mellon College of Science, 4400 Fifth Avenue, Pittsburgh, PA 15213, USA;
Email: \texttt{rwdoerge@andrew.cmu.edu}.} \ and Sanat K.
Sarkar\thanks{Department of Statistical Science and Fox School of Business,
Temple University, Philadelphia, PA 19122, USA; Email:
\texttt{sanat@temple.edu}.} }
\date{}
\maketitle

\begin{abstract}
Multiple testing with false discovery rate (FDR) control has been widely
conducted in the ``discrete paradigm" where p-values have discrete and
heterogeneous null distributions. However, in this scenario existing FDR procedures often lose
some power and may yield unreliable inference, and for this scenario there does not seem to be
an FDR procedure that partitions hypotheses into groups, employs data-adaptive weights and is non-asymptotically conservative. We propose a weighted FDR procedure
for multiple testing in the discrete paradigm that efficiently adapts to both
the heterogeneity and discreteness of p-value distributions. We theoretically
justify the non-asymptotic conservativeness of the weighted FDR procedure
under independence, and show via simulation studies that, for multiple testing
based on p-values of Binomial test or Fisher's exact test, it is more powerful
than six other procedures. The weighted FDR procedure is applied to a drug
safety study and a differential methylation study based on discrete data, where it makes more discoveries than two existing methods.

\end{abstract}


\section{Introduction}

\label{sec:Intro}

Multiple testing (MT) aiming at false discovery rate (FDR) control has been
routinely conducted in various scientific endeavours. With the deluge of
discrete data collected in genomics \citep{Auer:2010}, genetics
\citep{Gilbert:2005}, drug safety monitoring and other areas, researchers
frequently need to conduct multiple testing in the ``discrete paradigm'' where
test statistics have discrete and different distributions under null
hypotheses. Two major types of test conducted in the discrete paradigm are the
Binomial test (BT) and Fisher's exact test (FET). As reported by
\cite{Gilbert:2005} and \cite{Pounds:2006}, discreteness and heterogeneity of
p-value distributions under null hypotheses can reduce the power of two of the
most popular FDR procedures, i.e., the ``BH'' procedure in
\cite{Benjamini:1995} and ``Storey's procedure'' in \cite{Storey:2004}, that
were initially designed for the ``continuous paradigm'' where test statistics have a
continuous and identical distribution under null hypotheses. This has
triggered considerable efforts to improve existing FDR procedures.

To adjust existing procedures for discrete distributions, minimal achievable
significance levels \citep{Gilbert:2005,Heyse:2011}, less conservative
estimators of the proportion $\pi_{0}$ of true null hypotheses
\citep{Liang:2015,Pounds:2006,Chen:2016} or randomized p-values
\citep{Habiger:2015} have been used. On the other hand, to account for
heterogeneous null distributions, almost all methods assume a Bayesian
two-component mixture model or a variant of this model where a p-value under
the true null hypothesis is (conditionally) uniformly distributed on the compact interval $\left[
0,1\right]  $; see \cite{Cai:2009}
and \cite{Ignatiadis:2016}. Clearly, these models are inappropriate for
p-values of BTs or FETs. Even though \cite{Hu:2010} proposed the Group
Benjamini-Hochberg procedure (``GBH'') that is able accommodate heterogeneous
null distributions, they assumed that groups of hypotheses with different
proportions of true null hypotheses exist, and were only able to prove the
asymptotic conservativeness of GBH by assuming that various
empirical processes converge to the two-component mixture model mentioned
earlier. Finally, even though the work of \cite{Dohler:2017} is able to
accommodate discreteness and heterogeneity of p-value distributions, its
authors admitted that their methods may be more conservative than the BH
procedure (due to their construction of critical constants), and its implementation can be quite computationally intensive when
the number of hypotheses to test is large (due to the need to search over subsets of p-values in order to construct critical constants).

\subsection{Main contributions}

In this work, we propose a weighted FDR (``wFDR") procedure to better tackle
both the heterogeneity and discreteness of p-value distributions. The wFDR
procedure utilizes p-value grouping and weighting, does not employ any
Bayesian mixture model, does not assume the existence of groups of hypotheses
with different proportions of true null hypotheses, and is theoretically shown
to be conservative non-asymptotically regardless of how p-values are
partitioned into groups.

Specifically, to tackle heterogeneity, we propose a new grouping algorithm to partition hypotheses into
groups such that distributions of p-values within a group are less
heterogeneous, so that the wFDR procedure can better utilize information from p-value
distributions within a group. The algorithm can be easily and fast implemented
for distributions of p-values of BTs or FETs even when there are a large number of hypotheses to test and each p-value cumulative distribution function (CDF) has a large support. Once hypotheses are partitioned
into groups, the wFDR procedure employs data-adaptive weights constructed by \cite{Nandi:2018}, which are based on a different strategy than the plug-in strategy
of \cite{Hu:2010}. The groupwise weights account for the discreteness and remaining heterogeneity of p-value distributions for each group, and potentially utilize the power characteristics of the tests. Finally, the wFDR procedure applies the BH procedure
to the weighted p-values for multiple testing.

When p-values are super-uniform under null hypotheses and are
independent, we prove that the wFDR procedure is conservative non-asymptotically. Under the same setting,
we provide in \autoref{secPluginWfdr} a non-asymptotic upper bound on the FDR of the wFDR
procedure when the weights are constructed using the proportion estimator of
\cite{Chen:2016}. This bound also applies to the FDR of the adaptive GBH procedure of \cite{Hu:2010}, and is perhaps the best such bound for it.
A sufficient condition is given such that the wFDR procedure is more powerful than the ``adaptive BH procedure''
in \cite{Chen:2016}.
For MT based on p-values of BTs and FETs, we empirically show that the wFDR
procedure is much more powerful than this adaptive procedure, the ``randomized Storey procedure'' of \cite{Habiger:2015} and four other procedures,
and that the stability of the wFDR procedure is robust
to the number of groups that the hypotheses are partitioned into and the value of its tuning parameter.

An \textsf{R} package
``\href{https://cran.r-project.org/web/packages/fdrDiscreteNull/index.html}{fdrDiscreteNull}%
'' to implement the wFDR procedure and other procedures compared in this
article is available on \href{https://cran.r-project.org/}{CRAN}.

\subsection{Relation to existing weighted FDR procedures}

There are quite a few multiple testing procedures that are based on p-value
weighting or grouping and aim at conservative FDR control. However, they can
be roughly categorized into:

\begin{itemize}
\item Class I: methods that assume p-values follow the uniform distribution on $[0,1]$
 marginally or conditionally under the null hypothesis, some of
which have weights that are obtained from complicated optimization algorithms;
see, e.g.,
\cite{Benjamini1997,Genovese:2006,Finos:2007,Cai:2009,Roquain:2009,Ignatiadis:2016}%
.

\item Class II: methods that assume p-values marginally under the null
hypothesis are super-uniform and are not theoretically guaranteed to be
conservative non-asymptotically, some of which have weights that potentially
satisfy restrictive conditions; see \cite{Hu:2010,Li:2017}.

\item Class III: method that assume the existence of groups of hypotheses with
different proportions of true null hypotheses; see \cite{Cai:2009,Hu:2010}.

\item Class IV: methods that employ weights that are induced by a covariate and are assumed to be independent of their corresponding p-values under the null hypothesis,
or that employ an informative variable that is independent of p-value under the null hypotheis and a functional proportion of true null hypothesis;
see \cite{Ignatiadis:2016,Chen:2017ffdr,Ignatiadis:2018}.
\end{itemize}

In the discrete paradigm where p-values have discrete and heterogeneous null
distributions and where hypotheses do not necessarily form groups that have
different proportions of true null hypotheses, none of the methods in Class I to III
are applicable in terms of their modelling assumptions or are fully
reliable in terms of their non-asymptotic FDR conservativeness. Further, for methods in Class IV, an unresolved issue is to ensure that the induced weights are
independent of their corresponding p-values under the null hypothesis, or to ensure the independence between the informative variable and the p-value under the true null hypothesis. In contrast,
the wFDR procedure are free from the shortcomings of all methods mentioned above. Finally, unlike the procedures of \cite{Dohler:2017}, the wFDR procedure is often
more powerful than the BH procedure (as shown by our simulation studies) and does not need to search over subsets of p-values.

\subsection{Organization of article}

The rest of the article is organized as follows. \autoref{Sec:NewMethods}
presents the rationale for the wFDR procedure, new grouping algorithm, wFDR
procedure, and a theorem on the conservativeness of the wFDR
procedure. A simulation study is provided in \autoref{Sec:EmpiricalStudy}. Two applications of the wFDR procedure to
multiple testing with discrete data are provided in \autoref{Sec:AppReal}. The
article is concluded by \autoref{Sec:Discussion}. All proofs and the wFDR procedure with plug-in weights are provided in the appendices.

\section{A weighted FDR procedure for discrete paradigm}

\label{Sec:NewMethods}

In this section, we will illustrate the rationale behind the wFDR procedure
using multiple testing means of Poisson and Binomial random variables (see
\autoref{secSetting} and \autoref{sec:rationale}), and introduce the new
grouping algorithm (see \autoref{Sec:NewGrouping}) and the wFDR procedure with
its non-asymptotic conservativeness (see \autoref{Sec:NewGFDRProcedure}).

\subsection{Multiple testing Poisson means or Binomial means}

\label{secSetting}

In a typical multiple testing setting, there are $m$ null hypotheses to be
tested simultaneously, among which $m_{0}$ are true and the rest $m_{1}$ are
false. Further, associated with each hypothesis is a test statistic and its
p-value. The proportion $\pi_{0}$ of true null hypotheses is defined as
$\pi_{0} = m_{0} /m$, and a multiple testing procedure (MTP) is usually
applied to the $m$ p-values. Our rationale for the wFDR procedure
comes from analyzing multiple testing equality of Poisson means or Binomial
means, where the Binomial test (BT) or Fisher's exact test (FET) is used; see
\cite{Lehmann:2005} for details on these two types of test.

Multiple testing Poisson means can be described as follows. Let $\mathsf{Pois}%
\left(  \mu\right)  $ denote a Poisson random variable (or its distribution)
with mean $\mu$. Assume there are $2m$ mutually independent Poisson random
variables $\mathsf{Pois}\left(  \mu_{si}\right)  $ with $s=1,2$ and
$i=1,\ldots,m$, such that $\mathsf{Pois}\left(  \mu_{1i}\right)  $ and
$\mathsf{Pois}\left(  \mu_{2i}\right)  $ form a pair for each $i$. To access
how many pairs of these Poisson random variables have equal means, for each
$i$ the BT is conducted to assess the null $H_{i0}:\mu_{1i}=\mu_{2i}$ versus
the alternative $H_{i1}:\mu_{1i}\neq\mu_{2i}$, a two-sided p-value $p_{i}$ is
obtained, and then an MTP is applied to the $m$ p-values to determine which
null hypotheses are true.

Multiple testing Binomial means can be described similarly. Let $\mathsf{Bin}%
\left(  q,n\right)  $ denote the Binomial random variable (or its
distribution) with probability of success $q$ and number of trials $n$. Assume
there are $2m$ mutually independent Binomial random variables $\mathsf{Bin}%
\left(  q_{si},n_{si}\right)  $ with $s=1,2$ and $i=1,\ldots,m$, such that
$\mathsf{Bin}\left(  q_{1i},,n_{1i}\right)  $ and $\mathsf{Bin}\left(
q_{2i},,n_{2i}\right)  $ form a pair for each $i$. To access how many pairs of
the Binomial random variables have equal means, for each $i$ FET is conducted
to assess the null $H_{i0}:q_{1i}=q_{2i}$ versus the alternative
$H_{i1}:q_{1i}\neq q_{2i}$, a two-sided p-value $p_{i}$ is obtained, and then
an MTP is applied to the $m$ p-values to determine which null hypotheses are
true. The construction of an FET is illustrated in \autoref{table:FETA}.
\begin{table}[t]
\centering
\begin{tabular}
[c]{l|l|l|l}\hline
& Observed & Unobserved & Total\\\hline
$\mathsf{Bin}\left(  q_{1i}, n_{1i}\right)  $ & $c_{1i}$ & $n_{1i}-c_{1i}$ &
$n_{1i}$\\
$\mathsf{Bin}\left(  q_{2i}, n_{2i}\right)  $ & $c_{2i}$ & $n_{2i}-c_{2i}$ &
$n_{2i}$\\
Total & $M_{i}=c_{1i}+c_{2i}$ & $\tilde{M}_{i}=M_{i}^{\ast}-M_{i}$ &
$M_{i}^{\ast}=n_{1i}+n_{2i}$\\\hline
\end{tabular}
\caption[A 2-by-2 table for FET]{A 2-by-2 table that illustrates, for Fisher's
exact test (FET), how the joint marginal vector $\mathbf{N}_{i}^{\ast}=\left(
n_{1i},n_{2i},M_{i}\right)  $ is formed based on observed counts from two
Binomial distributions $\mathsf{Bin}\left(  q_{si},n_{si}\right)  $ for
$s=1,2$.}%
\label{table:FETA}%
\end{table}

\subsection{Rationale for the weighted FDR procedure}

\label{sec:rationale}

The rationale behind grouping hypotheses and weighting p-values are two fold
and explained as follows. The heterogeneity of p-value distributions when
conducting the BT or FET is due to conditioning on different observed total
counts or marginal counts. However, for two discrete p-value distributions
obtained from either test to be close to each other under the null hypotheses,
it is necessary that the observed total or marginal counts be close to each
other. In other words, similar p-value distributions represent statistical
evidence of similar strength against the null hypotheses. Therefore,
partitioning hypotheses into groups according to the similarity between their
associated p-value distributions helps gather statistical evidence of similar
strength, accounts for the heterogeneity of the p-value distributions, and may
lead to more powerful decision rules.

On the other hand, by the definitions of BT and FET, a conditional test
induces a ``conditional proportion" for each null hypothesis. For example,
given $m$ pairs of independent Poisson distributions such that $m_{0}$ pairs
have equal means, and let the $i$th null hypothesis be that the $i$th pair of
Poisson distributions have equal means. Then $\pi_{0} = m_{0}/m$. When BT is
used to test the $i$th null hypothesis based on the total of the observed
counts of the $i$th pair of Poisson random variables, a conditional proportion
$\tilde{\pi}_{0i}$, dependent on the total count, is induced for the $i$th
null hypothesis. The same reasoning applies when FET is conducted. In other
words, due to conditional testing, $\pi_{0}$ induces random realizations such
that the $i$th null hypothesis has its own conditional proportion $\tilde{\pi
}_{0i}$ of being true. Thus, multiple testing based on p-values of conditional
tests, even though being cast in the frequentist paradigm, has a strong
Bayesian interpretation, in that each null hypothesis has its own probability
of being true and the p-value distributions are heterogeneous.

Accordingly, the weighted FDR procedure uses the new grouping algorithm to
partition discrete p-values into groups so that p-values in each group have
less heterogeneous distributions, forms weights for p-values in each group
using the concept of Bayesian FDR \citep{Efron:2001}, and interpolates the corresponding Bayesian decision
rule; see \autoref{Sec:NewGFDRProcedure} for details on the weighted FDR procedure.

\subsection{A new grouping algorithm}

\label{Sec:NewGrouping}

For $i=1,...,m$, let $H_{i}$ denote the status of the $i$th null hypothesis,
such that $H_{i}=0$ or $1$ means that the $i$th null hypothesis is true or
false. Let $p_{i}$ be the p-value associated with the $i$th null hypothesis.
We take the convention that any CDF is
right-continuous with left-limits and let $\mathcal{D}$ be the set of CDFs.
$\mathcal{D}$ contains many types of CDFs that are commonly encountered in
multiple testing. For example, it contains the CDF of the standard uniform random variable,
the CDFs of p-values induced by permutation tests, and the CDFs of p-values of
BTs or FETs. The metric we take to group CDFs is the infinity norm for
functions. Namely, the distance between any two $F$ and $G$ in $\mathcal{D}$
is defined as
\begin{equation}
\label{1}\delta\left(  F,G\right)  = \left\Vert F-G\right\Vert _{\infty} =
\sup\nolimits_{x \in\mathbb{R}} \vert F(x) - G(x) \vert.
\end{equation}
Clearly, the distance is $0$ between any two CDFs of p-values that are uniformly
distributed with the same discontinuities. For example, this is true for the
CDFs of p-values under the null hypothesis in the continuous paradigm and for
any two p-value CDFs obtained from permutation test.

For a p-value, call its distribution under the null hypothesis the ``null
distribution" and that under the alternative hypothesis the ``alternative
distribution". Let $F_{i}$ be the null distribution of p-value $p_{i}$. Set
$\mathcal{C} =\left\{  F_{i}:i=1,\ldots,m\right\}  $. To partition
$\mathcal{C}$ into $l_{\ast}$ groups with minimal group size $g_{\ast}$ such
that each group contains less heterogenous CDFs, we propose a grouping
algorithm $\mathcal{G}_{\delta}$ as \autoref{Alg1}. The grouping algorithm
accounts for the practical need of a minimal group size. Once the set
$\mathcal{C}$ is partitioned, the set of null hypotheses and that of p-values
are partitioned accordingly. When the null distributions of p-values are
identical such as those in the continuous paradigm or induced by permutation
test, $\delta\equiv0$ and $\mathcal{G}_{\delta}$ automatically forms one
group. Further, we do not assume that a group structure exists among
hypotheses.

Two attractive features of the algorithm are: (1) for a BT, its null
distribution is determined by the observed total count from the two Poisson
random variables, and therefore grouping CDFs of p-values of BTs can be
equivalently based on the Euclidean norm on the observed total counts
without computing the supremum norm; (2) for an FET (illustrated in
\autoref{table:FETA}), its null distribution is determined by the joint
marginal vector $\mathbf{N}_{i}^{\ast}=\left(  n_{1i},n_{2i},M_{i}\right)  $
from the two Binomial random variables, and therefore grouping CDFs of
p-values of FETs can be approximately based on the Euclidean
norm on the observed total counts $M_{i}$ if the Binomial variables have the
similar number of trials without computing the supremum norm. This will greatly speed up
computing the distances between discrete CDFs in order to form groups of hypotheses for BTs and FETs.
Thus, the grouping scheme used in the simulation study in
\autoref{Sec:EmpiricalStudy} approximately equivalently implements \autoref{Alg1}.

\begin{algorithm}[t]
{With $\mathcal{C} =\left\{  F_{i}:i=1,\ldots,m\right\}$, compute $\delta_{ij}=$ $\delta\left(
F_{i},F_{j}\right)  $ for all $1 \leq i \leq j \leq m$ and set $\delta^{\ast}%
=\max\nolimits_{1\leq i<j\leq m}\delta_{ij}$\;}
{{\bf input}: $\sigma=\left(  2l_{\ast}\right)  ^{-1}\delta^{\ast}$, $\tilde{l}=1$, $\tilde{g}_{\ast}=1$,
$B_{0}=\left\{  1,\ldots,m\right\}  $, $A_{0}=D_{0}=\varnothing$\;}
\While{$\tilde{g}_{\ast}<g_{\ast}$}{
\While {$\left\vert B_{0}\right\vert >0$ and $\tilde{l}\leq l_{\ast}$ }{
\For  {$i\in B_{0}$} {
{$B_{i}\left(\sigma\right) \gets \left\{  F_{j}:j\in B_{0},\delta_{ij}\leq\sigma\right\}$,
$A_{i} \gets \left\{  j\in B_{0}:F_{j}\in B_{i}\left(  \sigma\right)  \right\}$}\;
} 
{$i^{\ast} \gets \arg\max\left\{  i\in B_{0}:\left\vert A_{i}\right\vert \right\}  $,
$D_{0} \gets D_{0}\cup\left\{\left\{  A_{i^{\ast}}\right\}\right\}  $,
$B_{0} \gets B_{0}\backslash A_{i^{\ast}}$\;} 
\If {$\tilde{l} \leq l_{\ast}-1$}{
\If {$\left\vert B_{0}\right\vert \leq h_{\ast}$} {
set $A_{0},B_{0},D_{0},\tilde{l}$ as in {\bf input} but $\sigma \gets 0.5\sigma$\;}
\If {$\tilde{l} =l_{\ast}-1$ and $\left\vert B_{0}\right\vert =h_{\ast}$} {
$A_{l_{\ast}} \gets B_{0}$, $D_{0} \gets D_{0}\cup\left\{\left\{  A_{l^{\ast}}\right\}\right\}  $,
$\left\vert B_{0}\right\vert \gets 0$\;}
\If {$\left\vert B_{0}\right\vert >h_{\ast}$}{keep current $B_{0}$ and $D_{0}$\;}
} 
\If {$\tilde{l} =l_{\ast}$}{
\If {$\left\vert B_{0}\right\vert \leq h_{\ast}$}{$A_{l_{\ast}} \gets A_{l_{\ast}}\cup B_{0}$\;}
\If {$\left\vert B_{0}\right\vert >h_{\ast}$} {set $A_{0},B_{0},D_{0},\tilde{l}$ as in {\bf input} but
$\sigma \gets 1.5\sigma$\;}
} 
{$\tilde{l} \gets \tilde{l}+1$\;}
} 
{$\tilde{g}_{\ast} \gets \min_{1 \le j \le l_{\ast}}\left\vert A_{j}\right\vert$\;}
} 
\caption{Grouping algorithm to partition p-value CDFs into $l_{\ast}$ groups with minimal group size $g_{\ast}$ and merging size $h_{\ast}$}
\label{Alg1}
\end{algorithm}

The grouping \autoref{Alg1} basically works as follows. First the distance
$\delta_{ij}$ between each pair of $F_{i}$ and $F_{j}$ with $i \neq j$ is
computed, and the maximum $\delta^{\ast}$ of $\delta_{ij}$'s obtained. Note
that $\delta_{ii}=0$. Pick an initial radius $\sigma=\left(  2l_{\ast}\right)
^{-1}\delta^{\ast}$. Secondly,
for each $F_{i} \in\mathcal{C}$, form the subset $B_{i}\left(  \sigma\right)
$ of CDFs such that their distances from $F_i$ are upper bounded by
$\sigma$; find among subsets $B_{i}\left(  \sigma\right)  , i=1,\ldots,m$ one,
denoted by $B_{{i}^{\ast}}\left(  \sigma\right)  $, that has the most members,
and remove it from $\mathcal{C}$. We call this ``identification and removal
process (IRP)". Note that $B_{{i}^{\ast}}\left(  \sigma\right)  $ is a
candidate group we have formed. Apply the IRP to the rest of the CDFs, and
repeat it by adjusting the value for $\sigma$ until $l_{\ast}$ groups
are obtained and no group has less than $g_{\ast}$ members. We remark that implementing $\mathcal{G}_{\delta}$
according to \autoref{Alg1} and computing supremum norms can take some time when $m$ is large and each $F_i$ has a large support.

In \autoref{Alg1}, whenever the current radius $\sigma$ in ``\textbf{input}''
is changed in the inner ``\textbf{while}'' loop into some $\sigma^{\prime} $,
we say that a new iteration inside $\mathcal{G}_{\delta} $ is started with the
new radius $\sigma^{\prime}$. The merging size $h_{\ast}$ is set to be no less
than the minimal group size $g_{\ast}$ to speed up grouping and ensure that
$\mathcal{G}_{\delta}$ terminates in a finite number of iterations. However,
the final radius the algorithm assumes to successfully form $l_{\ast}$ groups
with minimal group size $g_{\ast}$ may be different than the initial radius
$\sigma$. For notational simplicity, we will still use $\sigma$ to denote the
final radius. Since $\delta$ is a metric on the Cartesian product $\mathcal{C}
\otimes\mathcal{C} $, the grouping algorithm $\mathcal{G}_{\delta}$ always
produces out of $\mathcal{C}$ at least $l_{\ast}-1$ $\delta$-metric balls of radii upper bounded by $\sigma$.
$\mathcal{G}_{\delta}$ may form the $l_{\ast}$th group by
merging the last group with at most $h_{\ast}$ members or by taking the last
$h_{\ast}$ members; see the first subcase of the case $\tilde{l}=l_{\ast}$ or
the second subcase of the case $\tilde{l}=l_{\ast}-1$ in \autoref{Alg1}. In
this case, the $l_{\ast}$th group may contain p-value CDFs that are more than
$2\sigma$ units away in the $\delta$-metric.

For each natural number $s$, let $\mathbb{N}_{s}=\left\{1,\ldots,s\right\}  $, and for a set $A$,
let $\left\vert A\right\vert $ be its cardinality and $\mathbf{1}_{A}$ its indicator. The grouping algorithm
$\mathcal{G}_{\delta}$ retains maximal within-group homogeneity in the sense of the following result:

\begin{theorem}\label{groupAlgo}
  Let $\mathbb{G}$ be the set of partitions of $\mathcal{C} =\left\{  F_{i}:i=1,\ldots,m\right\}$ into $l_{\ast} \ge 2$ groups that consist of at least $l_{\ast}-1$ $\delta$-metric balls of radii no larger than $\sigma$.
  Let $\tilde{G}_j$ be the $j$th group produced by $\mathcal{G}_{\delta}$ and $n_{j} = \vert \tilde{G}_j \vert$ for each $1 \le j \le l_{\ast}$, and let $\{n_{j}^{\prime}\}_{j=1}^{l_{\ast}-1}$ be the cardinalities of any set of $l_{\ast}-1$ $\delta$-metric balls of radii no larger than $\sigma$ produced by any $\mathcal{K} \in \mathbb{G}$.
  With loss of generality, assume $n_{j}^{\prime}$'s are ordered descendingly in $j$ for $1\leq j\leq l_{\ast}-1$.
Then $n_{j}\geq n_{j}^{\prime}$ for $1\leq j\leq l_{\ast}-1$.
\end{theorem}
\noindent
The proof of \autoref{groupAlgo} follows directly from the definition of $\mathcal{G}_{\delta}$ and is thus omitted.
Roughly speaking, \autoref{groupAlgo} says that among all partitions of size $l_{\ast} \ge 2$ of the set $\mathcal{C}$ of CDFs into at least $l_{\ast}-1$ $\delta$-metric balls of radii no larger than $\sigma$, each of the first $l_{\ast} -1$ groups produced by $\mathcal{G}_{\delta}$ contains the maximal number of CDFs and thus achieves maximal within-group homogeneity in terms of distributional differences.

\subsection{The weighted FDR procedure and its conservativeness}

\label{Sec:NewGFDRProcedure}

Let $I_{0}$ be the index set of true nulls with cardinality
$m_{0}$, and $\pi_{0}=m_{0}m^{-1}$ the proportion of true nulls.
We assume throughout the rest of the article that $\min\left\{  p_{i}%
:i\geq1\right\}  >0$ almost surely, in order to avoid the undetermined
operation $0\times\infty$ when a p-value is $0$ and a weight is $\infty$. The wFDR procedure is stated as follows:

\begin{itemize}
\item \textbf{Grouping}: apply the grouping algorithm $\mathcal{G}_{\delta}$
to partition $\mathcal{C}=\left\{  F_{i}:i=1,\ldots,m\right\}  $ into groups
$\tilde{G}_{j},j=1,\ldots,l_{\ast}$, whose corresponding groups $G_{j}%
,j=1,\ldots,l_{\ast}$ of indices partition the index set $\mathbb{N}_{m}=\left\{1,\ldots,m\right\}$.

\item \textbf{Constructing weights}: fix a $\lambda\in\left(  0,1\right)  $, the tuning parameter, and for each $j$ and $G_{j}$, set%
\begin{equation}
w_{j}=\frac{\left(  n_{j}-R_{j}\left(\lambda\right) +1\right)  \left( R \left(\lambda\right) +l_{\ast}-1\right)  }{m\left(  1-\lambda
\right)  R_{j}\left(\lambda\right) },\label{eqe2a}%
\end{equation}
where $R_{j}\left(\lambda\right)= \sum_{i\in G_{j}}\mathbf{1}_{\left\{  p_{i}\leq\lambda
\right\}} $, $R \left(\lambda\right)= \sum_{i=1}^{m}\mathbf{1}%
_{\left\{  p_{i}\leq\lambda\right\}  }$ and $n_j = \vert G_j \vert$.

\item \textbf{Weighting and rejecting}: weight the p-values $p_{i}$, $i\in
G_{j}$ into $\tilde{p}_{i}=p_{i}w_{j}$, and apply the BH procedure to
$\left\{  \tilde{p}_{i}\right\}  _{i=1}^{m}$.
\end{itemize}

Note that $w_j= m^{-1}\left(  1-\lambda\right)^{-1}\left(  m-R\left(\lambda\right) +1\right)$ when $l_{\ast}=1$, and that $a/0=\infty$ is set for \eqref{eqe2a} when $a > 0$.
The weights $w_j,j=1,\ldots,l_{\ast}$ were constructed by \cite{Nandi:2018} and have been adopted here. We refer the readers to \cite{Nandi:2018} for the intuition behind these weights, their connections with the proportion estimator of \cite{Storey:2004}, and their differences from the plug-in weights of \cite{Hu:2010}. Specifically, $w_j$ is the product of two estimates
$n_j^{-1}\left(  1-\lambda\right)^{-1}\left(  n_{j}-R_{j}\left(\lambda\right) +1\right)$ and $n_jR_{j}^{-1}\left(\lambda\right) m^{-1}\left( R \left(\lambda\right) +l_{\ast}-1\right)$
respectively of $\pi_{j0}^{\ast}$ and $ (1-\pi_{j0}^{\ast})^{-1} \left(  1-\pi_{0}\right)$, where $\pi_{j0}^{\ast}$ is the proportion of true null hypotheses among the $j$-th group of hypotheses $\mathcal{H}_{j}=\left\{  H_{j_{k}}:k\in
G_{j}\right\}  $. Note that $\pi_{j0}^{\ast} (1-\pi_{j0}^{\ast})^{-1} \left(  1-\pi_{0}\right)$ is the weight for group $j$ for the oracle GBH of \cite{Hu:2010}.

We refer to as \textquotedblleft null p-value\textquotedblright\ a p-value
whose associated null hypothesis is true and as \textquotedblleft alternative
p-value\textquotedblright\ a p-value whose associated null hypothesis is
false. Further, we refer to a p-value as
\textquotedblleft super-uniform\textquotedblright\ if its CDF $F$
satisfies $F(t)\leq t$ for all $t\in\lbrack0,1]$.
The following result justifies the non-asymptotic conservativeness of the wFDR procedure, and provides a condition under which the wFDR
procedure non-asymptotically has no less rejections than the ``aBH procedure'' of \cite{Chen:2016}.

\begin{theorem}\label{ThmWFDR}
   Let $\hat{\pi}_{0}^{\left(  1\right)  }$ be the estimator of $\pi_{0}$ that is employed by the aBH procedure. Then on the event that
\begin{equation}
  \hat{\pi}_{0}^{\left(  1\right)  } \ge \min_{1\leq j\leq l_{\ast}}w_{j},\label{condweights}%
\end{equation}
the wFDR\ procedure rejects as least as many null hypotheses as the aBH procedure. On the other hand, if
 $\left\{p_i\right\}_{i=1}^{m} $ are independent and the null p-values are super-uniform, then the wFDR procedure is conservative.
\end{theorem}

\autoref{ThmWFDR} asserts that the wFDR procedure is conservative non-asymptotically regardless of how the null hypotheses are partitioned and the feasible value of the tuning parameter. We remark that the new algorithm $\mathcal{G}_{\delta}$ is employed to enhance the power of the weighted procedure when p-values have heterogeneous null distributions. In fact, \autoref{ThmWFDR} holds when $\mathcal{G}_{\delta}$ is replaced by any partition of the null hypotheses, which can be seen from its proof. \autoref{ThmWFDR} extends
the conservativeness of the adaptive one-way BH in \cite{Nandi:2018} from multiple testing in the continuous paradigm to the discrete paradigm.
Condition (\ref{condweights}) in \autoref{ThmWFDR} is simple, easy to check but a bit restrictive. Relaxing (\ref{condweights}) involves a careful study on the orderings of the weighted p-values induced by the two procedures compared by \autoref{ThmWFDR}. This is a very complicated task when the number of hypotheses to test is large, which we will not pursue here.

\section{Simulation study}

\label{Sec:EmpiricalStudy}

We present a simulation study on multiple testing based on p-values of Binomial tests
(BTs) and Fisher's exact tests (FETs) to compare the wFDR procedure (``wFDR'')
with six other procedures: the BH procedure (``BH''), the adaptive BH procedure (``aBH") of \cite{Chen:2016}, the procedure (``BHH'') of \cite{Heyse:2011}, the randomized Storey procedure (``SR'') of \cite{Habiger:2015}, the ``aHSU'' procedure of \cite{Dohler:2017}, and the ``BH+'' procedure of \cite{Chen:2018d}. The BH+ procedure is proven to be conservative and will be applied to two-sided mid p-values and be denoted by ``BH+MidP''; a definition of mid p-value can be found in \cite{Hwang:2001}. For the
simulated data, there do not exist groups of hypotheses for which each group
has its own proportion of true null hypotheses, and condition (\ref{condweights}) in \autoref{ThmWFDR}
is not necessarily satisfied.

\subsection{Simulation design under independence}

\label{subsec:SimSet}

For $a < b$, let $\mathsf{Unif}\left(  a,b\right)  $ be the uniform random
variable or the uniform distribution on the closed interval $[a,b]$. The
simulations are set up as follows:

\begin{enumerate}
\item Set the number of hypotheses $m=5000$, the proportion of true null
hypotheses $\pi_{0} \in\left\{  0.5,0.6,0.7,0.8,0.95\right\}  $, nominal FDR level
$\alpha\in\left\{  0.05,0.1,0.15,0.2\right\}  $, the number of groups for the
weighted FDR procedure to partition the hypotheses into $l_{\ast} \in\left\{
3, 7, 10\right\}  $, the number of true null hypotheses $m_{0}=\left[
m\pi_{0}\right]  $, and $m_{1}=m-m_{0}$, where $\left[  x\right]  \ $is the
integer part of a real number $x$.

\item Two types of discrete data are generated as follows:

\begin{enumerate}
\item Poisson data: generate means $\mu_{1i}, i=1,\ldots,m$ from the Pareto
distribution $\mathsf{Pareto}\left(  7,7\right)  $ with scale parameter $7$
and shape parameter $7$. Generate $m_1$ $\rho_{i}$'s
independently from $\mathsf{Unif}\left(  1.5,5\right)  $. Set $\mu_{2i}=\mu_{1i}$ for $i=1,\ldots,m_{0}$,
$\mu_{2i}=\mu_{1i}\rho_{i}^{-1}$ for $m_{0}+1\leq i\leq
m_{0}+\left[  0.5m_{1}\right]  $, and $\mu_{2i}=\rho_{i}\mu_{1i}$ for $m_{0}+\left[  0.5m_{1}\right]  +1\leq i\leq m$.
For each $s$ and $i$, generate a count $c_{si}$ from the Poisson distribution
$\mathsf{Pois}\left(  \mu_{si}\right)  $ with mean $\mu_{si}$. For each $i$,
conduct the BT to assess the null $H_{i0}:\mu_{1i}=\mu_{2i}$ versus the
alternative $H_{i1}:\mu_{1i} \neq\mu_{2i}$ and obtain two-sided p-value
$p_{i}$.

\item Binomial data: generate $q_{1i}$ from $\mathsf{Unif}\left(
0.02,0.15\right)  $ for $i =1,\ldots,m_{0}$ and set $q_{2i}=q_{1i}$ for $i
=1,\ldots,m_{0}$. Set $q_{1i}=0.3$ and $q_{2i}=0.15$ for $m_{0}+1\leq i\leq
m_{0}+\left[  0.5m_{1}\right]  $, and $q_{1i}=0.15$ and $q_{2i}=0.3$ for $m_{0}+\left[  0.5m_{1}\right]  +1\leq i\leq m$.
Set $n=50$, and for each $s$ and $i$ generate a count $c_{si}$ from the
Binomial distribution $\mathsf{Bin}\left(  q_{si}, n\right)  $ with
probability of success $q_{si}$ and number of trials $n$. For each $i$, form a
2-by-2 table as \autoref{table:FETA} for which the observed counts are
$\left\{  c_{si}\right\}  _{s=1}^{2}$ and the marginal vector is
$\mathbf{N}_{i}^{\ast}=\left(  n,n,c_{1i}+c_{2i}\right)  $; apply FET to test
the null $H_{i0}:q_{1i}=q_{2i}$ versus the alternative $H_{i1}:q_{1i} \neq
q_{2i}$ and obtain two-sided p-value $p_{i}$.
\end{enumerate}

\item Apply the competing FDR procedures and record statistics for the
performance of each procedure. In order to implement the wFDR procedure, let
$\hat{c}_{i}=c_{1i}+c_{2i}$ for each $i$, set $q_{j},j=0,\ldots,l_{\ast}$ as
the $100 j l_{\ast}^{-1}$-th percentile of $\left\{  \hat{c}_{i}\right\}
_{i=1}^{m}$, and partition the hypotheses into $l_{\ast}$ groups, for which
group $G_{j} = \left\{  1 \leq i \leq m: q_{j-1}\leq\hat{c}_{i} <
q_{j}\right\}  $ for $1 \leq j \leq l_{\ast}-1$ and $G_{l_{\ast}} = \left\{  1
\leq i \leq m: q_{l_{\ast}-1}\leq\hat{c}_{i} \leq q_{l_{\ast}}\right\}  $.
This partitioning scheme is based on the Euclidian distance between the
observed total counts for both types of tests, as discussed in
\autoref{Sec:NewGrouping}.

\item For each combination of the triple $\pi_{0}$, $\alpha$ and $l_{\ast}$,
repeat Steps 2. and 3. $300$ times.
\end{enumerate}

For each test, its two-sided p-value is computed according to the formula in
\cite{Agresti:2002}, i.e., it is the probability computed under the null
hypothesis of observing values of the test statistic that are equally likely
to or less likely than the observed test statistic.

\subsection{Simulation design under dependence}
\label{secSimDepsupp}

\renewcommand\thefigure{\thesection.\arabic{figure}}
\setcounter{figure}{0} \renewcommand\thetable{\thesection.\arabic{table}} \setcounter{table}{0}

Set $m=5000$. The simulation design to generate dependent Binomial and Poisson
data is as follows:

\begin{enumerate}
\item Construct a block diagonal, correlation matrix $\mathbf{D}$ with $100$ blocks of equal sizes as follows: for the first $20$ blocks each block has identical off-diagonal entries $0.1$, for the second $20$ blocks each block as identical off-diagonal entries $0.2$, ..., and for the fifth $20$ blocks each block as identical off-diagonal entries $0.5$. Namely, each block of
$\mathbf{D}$ corresponds to a random vector whose entries are equally
correlated. Note that there is no specific reason for choosing $100$ blocks for such a $\mathbf{D}$.

\item Generate a realization $\mathbf{z}=(z_{1},\ldots,z_{m})$ from the
$m$-dimensional Normal distribution with zero mean and correlation matrix
$\mathbf{D}$, and obtain the vector $\mathbf{u}=(u_{1},\ldots,u_{m})$ such
that $u_{i} = \Phi(z_{i})$, where $\Phi$ is the CDF of the standard Normal
random variable. By its definition, $u_{i}, i=1,\ldots,m$ have a block
dependence structure inherited from $\mathbf{D}$.

\item Maintain the parameters and other settings used in \autoref{subsec:SimSet}; generate $m$-dimensional vectors of Binomial or Poisson random
variables using the vector $\mathbf{u}$ as the quantiles of the corresponding
marginal Binomial or Poisson distribution. Namely, for each $s \in\left\{
1,2\right\}  $ and $i \in\left\{  1,\ldots,m\right\}  $, the generated count
$c_{si}$ corresponds to quantile $u_{i}$ of the CDF of $\mathsf{Pois}\left(
\mu_{si}\right)  $ or $\mathsf{Bin}\left(  q_{si}, n\right)  $.
\end{enumerate}

We intend to use the above design to generate vectors of Binomial or Poisson
random variables whose correlation matrices are block diagonal and
approximately positive, since this type of dependence has been a popular
touchstone to assess the performance of an FDR procedure under dependence. For this simulation setting, we will
only present the performance of the weighted FDR procedure when it partitions
the hypotheses into $3$ groups and employs tuning parameter $\lambda=0.5$.

\subsection{Simulation results}

\label{Sec:SimRes}

We use the expectation of the true discovery proportion (TDP), defined as the
ratio of the number of rejected false null hypotheses to the total number of
false null hypotheses, to measure the power of an FDR procedure. We also
report the standard deviation of the FDP of a procedure since a smaller standard
deviation implies that the corresponding procedure is more stable in its FDR. Note that FDR is the expectation of FDP. Since
the aHSU procedure only has software implementation for multiple testing based on p-values of FETs, it was not implemented for the scenario of multiple testing based on p-values of BTs.
However, to visualize the simulation results via \textsf{facet} in the \textsf{R} software \textsf{ggplot2}, we set the FDP and TDP of the aHSU procedure to be zero for this
scenario.

\begin{figure}[htp]
\centering
\includegraphics[height=0.7\textheight,width=0.9\textwidth]{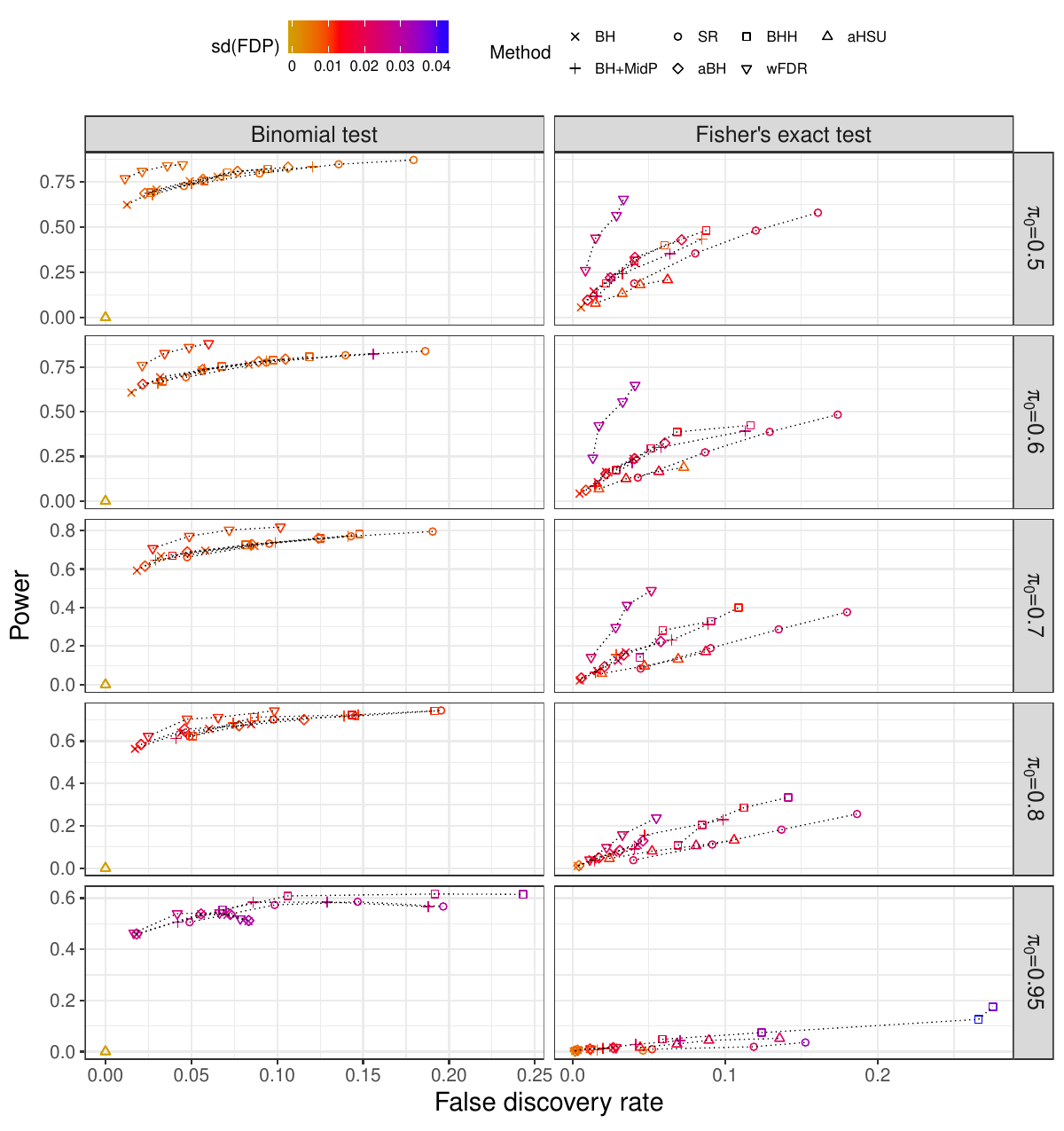}
\caption[FDR and power:two-sided p-values; 3 groups]{FDR and power of each procedure, where the weighted FDR procedure (``wFDR") partitions the
hypotheses into $3$ groups and employs tuning parameter $\lambda=0.5$. In the legend, ``sd(FDP)'' is the estimated standard deviation of the FDP of a procedure. Each type of points from left to
right in each subfigure are obtained successively under nominal FDR levels
$0.05,0.1,0.15, 0.2$. The weighted FDR procedure (``wFDR") is the most powerful, and the smaller
$\pi_{0}$ is, the more improvement in power the wFDR procedure has over the
other procedures.}%
\label{figfdrpwrFet}%
\end{figure}

First, we discuss the performances of the procedures under independence.
\autoref{figfdrpwrFet} presents the FDRs and powers of the procedures, for which wFDR has partitioned the hypotheses
into $3$ groups. The simulation results show that: (i) All procedures but BHH are conservative. This is reasonable since BHH has not been proven to be conservative.
All procedures are stable, i.e., having small standard
deviations for the FDPs. (ii) wFDR is the most powerful, BHH the second, SR the third, BH+MidP  the fourth, aBH the third, and BH and aHSU
the least. (iii) The smaller the proportion $\pi_{0}$ is, the more improvement
in power wFDR has over the other procedures.

\begin{figure}[htp]
\centering
\includegraphics[height=0.7\textheight,width=0.9\textwidth]{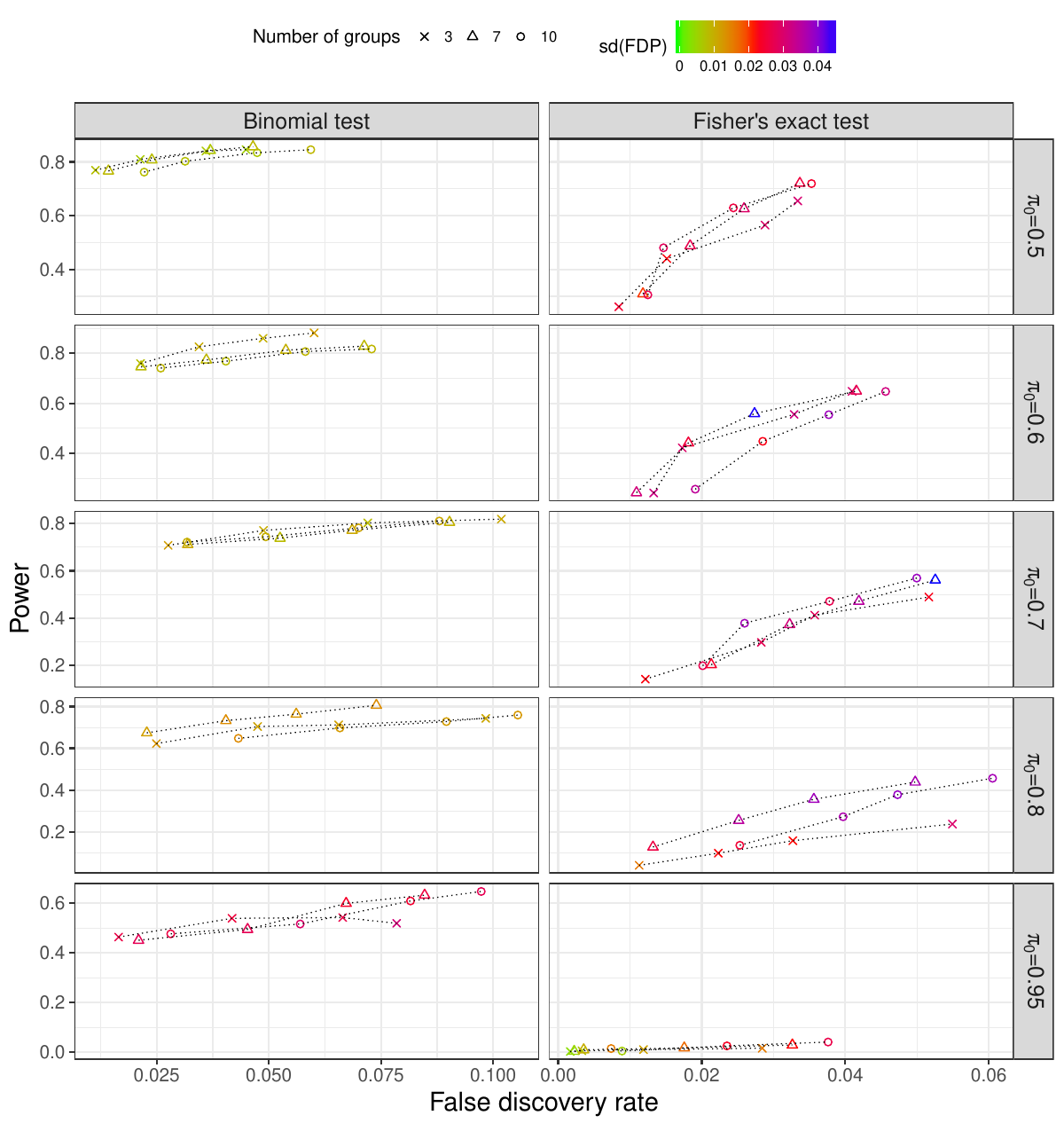}
\caption[wFDR: two-sided p-values; all groups]{FDR and power of the weighted FDR
procedure when it partitions the hypotheses into different ``Number of groups'' and tuning parameter $\lambda=0.5$.
In the legend, ``sd(FDP)'' is the estimated standard deviation of the FDP of a procedure. Each type of points from left to
right in each subfigrue are obtained successively under nominal FDR levels
$0.05,0.1,0.15, 0.2$. We see that increasing the number of groups to partition
the hypotheses into can increase the power of the wFDR procedure. However,
beyond a value for the number of groups, the power of the wFDR procedure stops
increasing and may decrease.}%
\label{figNew}%
\end{figure}

\begin{figure}[htp]
\centering
\includegraphics[height=0.75\textheight,width=.9\textwidth]{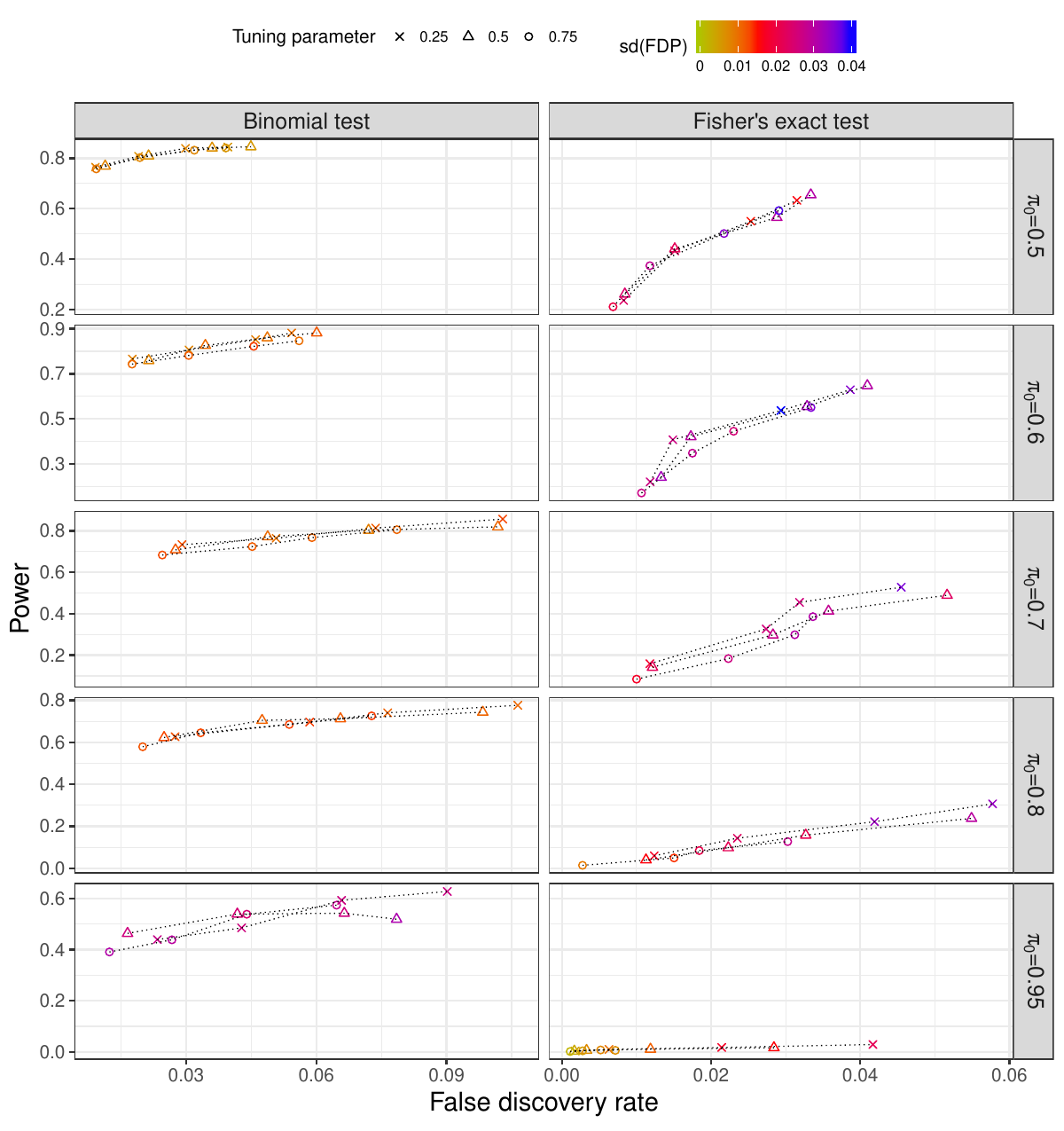}
\caption[FDR and power: two-sided; 3 lambda values]{FDR and power of the weighted FDR procedure when it partitions the
hypotheses into $3$ groups but employs different values of ``Tuning parameter'' $\lambda$.
In the legend, ``sd(FDP)'' is the estimated standard deviation of the FDP of a procedure. Each type of points from left to
right in each subfigure are obtained successively under nominal FDR levels
$0.05,0.1,0.15, 0.2$. We see that changing the value of $\lambda$ does not considerably affect the FDR or power
of the wFDR procedure.}%
\label{figpwr7}%
\end{figure}

For wFDR, \autoref{figNew} and \autoref{figpwr7} respectively present its FDR and power, as the number $l_{\ast}$
of groups the procedure partitions the hypotheses into ranges in $\{3,7,10\}$ and as the tuning parameter $\lambda$ ranges in
$\{0.25,0.5,0.75\}$.
These figures show the following: (i) Increasing the value of $l_{\ast}$ usually does not decrease the power or
increase the FDR of wFDR, or increase the standard deviation of the FDP or TDP of wFDR. (ii) The power of wFDR may stop
increasing and its FDR may stop decreasing once $l_{\ast}$ is large, e.g.,
when $l_{\ast} >7$ for the simulations. This is reasonable since the data adaptive-weights will likely increase with $l_{\ast}$ when $l_{\ast}$ is large and beyond a specific value (that is determined by the distributional properties of the p-values), thus reducing the power of wFDR. (iii) Changing the value of $\lambda$ does not considerably affect the FDR, power or stability of wFDR.

The performances of wFDR when $\lambda=0.5$ and $l_{\ast}=7$ or $10$ respectively are compared to those of the other six procedures. In these settings, wFDR is stable and is still the most powerful. Further, with $l_{\ast} =3$, we assess the power of wFDR when the null hypotheses are partitioned by k-means (applied to the 2-vectors each of which consists of the difference between the two observed counts and the total of the two observed counts for each hypothesis), and found that wFDR based on this partitioning method is less powerful than it is based on quantiles of the observed total counts (stated in the simulation design in \autoref{subsec:SimSet}). To save space, the associated visualizations are not included in the paper.

\begin{figure}[htp]
\centering
\includegraphics[height=0.7\textheight,width=0.9\textwidth]{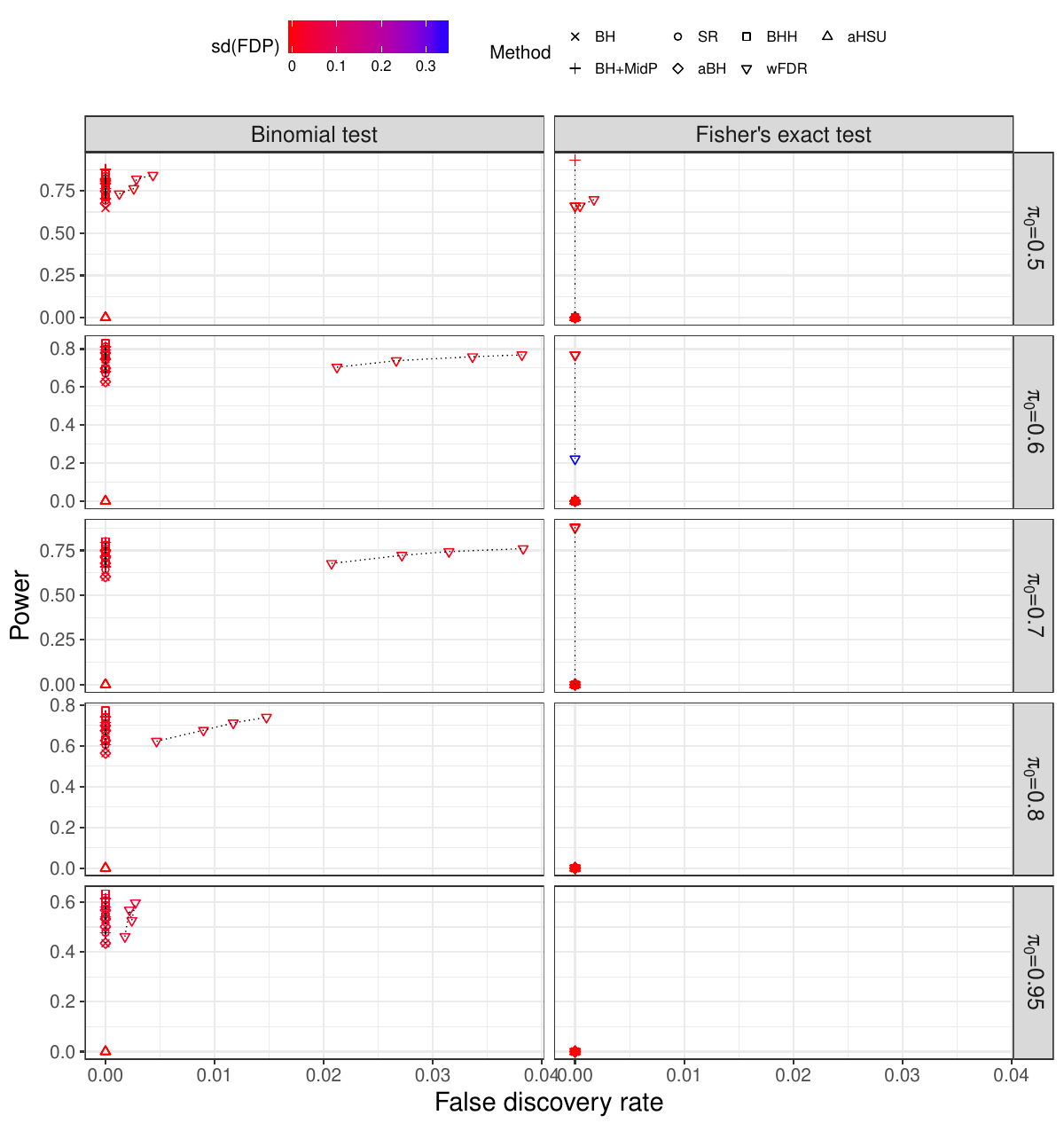}
\caption[FDR and power: two-sided p-values]{FDR and power of each FDR procedure
when it is applied to two-sided p-values and when the weighted FDR procedure (``wFDR") partitions the hypotheses into 3 groups and employs tuning parameter $\lambda=0.5$. In the legend, ``sd(FDP)'' is the estimated standard deviation of the FDP of a procedure. Each type of
points from left to right in each subfigure are obtained successively under
nominal FDR levels $0.05,0.1,0.15, 0.2$. We see that wFDR is competitive in power.
}%
\label{figfdrpwrFetdep}%
\end{figure}

Finally, we discuss the performance of the wFDR procedure under (approximate) positive dependence (see the simulation design in \autoref{secSimDepsupp}). Firstly, there is strong evidence that wFDR is conservative and competitive in power, regardless of when it is applied to p-values of BTs or FETs.
Secondly, we have observed that each FDR procedure in comparison can have
counter-intuitively large power and very small FDR when the nominal FDR level
increases and $\pi_{0}$ is not very close to $1$; see \autoref{figfdrpwrFetdep}. Such behavior
of the procedures for dependent discrete p-values
does not seem to have been reported elsewhere before. A partial explanation for this may be the clustering of signals
associated with the false null hypotheses and that of noise associated with the true null hypotheses, both due to dependence,
so that the wFDR procedure can have very low or high power depending on if its random rejection threshold badly or well splits the null and alternative p-values.
However, the exact reason deserves further investigation since in general we do not have a good understanding of the behavior of the random rejection threshold
of a multiple testing procedure or the empirical distribution of adaptively weighted p-values under dependence.

\clearpage
\section{Two applications to multiple testing with discrete data}

\label{Sec:AppReal}

We provide two applications of the wFDR procedure to multiple testing based on
discrete data, one for drug safety study and the other differential
methylation study, where p-values have discrete and heterogeneous null
distributions. The wFDR procedure will be implemented with $\lambda=0.5$ (based on our experience with
multiple testing with discrete p-values) and the grouping strategy in Step 3 of the
simulation design in \autoref{subsec:SimSet}, and it will be compared with the BH procedure
and ``aBH'' procedure of \cite{Chen:2016}. The proportion of true null hypotheses in each group and that for all hypotheses will be estimated by the estimator of \cite{Chen:2016}.

\subsection{Application to drug safety study}

The drug safety study aims to assess if a drug is associated with
amnesia. The data set, available from \cite{Heller:2012}, records the number
of amnesia cases and the total number of adverse events for each of the 2466
drugs. Among the total of 686911 reported adverse events, 2051 involves
amnesia cases. For each drug, the null hypothesis is that the drug is not
associated with amnesia, and there are $2466$ null hypotheses to test
simultaneously. The null hypothesis for drug $i$ will be tested by FET based
on \autoref{tdrug}. For this study, it is known that the number of drugs that
are associated with amnesia is small, i.e., the proportion $\pi_{0}$ of true
null hypotheses is close to $1$.

\begin{table}[ht]
\centering
\begin{tabular}
[c]{c|cc}
\hline
& Amnesia & Not Amnesia\\
\hline
Drug $i$ & $n_{1i}$ & $n_{2i}$\\
Other drugs & $2051-n_{1i}$ & $686911-2051-n_{2i}$\\
\hline
\end{tabular}
\caption{A 2-by-2 table used by Fisher's exact test (FET) to test if drug $i$
is associated with amnesia: $n_{1i}$ is the number of amnesia cases reported
for drug $i$, and $n_{1i}+n_{2i}$ the number of cases reported to have adverse
events for drug $i$. }%
\label{tdrug}%
\end{table}

This data set has been analyzed by \cite{Heller:2012} using one-sided p-values
of FETs and their FDR procedures. Here we re-analyze it using two-sided
p-values of FETs and the wFDR procedure. From \cite{chavant2011}, we know that
Benzodiazepines and anticholinergic drugs are often responsible for amnesia.
Based on this information, the wFDR procedure partitions the hypotheses into
$3$ groups, reflecting the side effects of the two known types of drugs and
that of the rest in the data.

The groupwise proportions are estimated to be $1$, $1$ and $0.98650$, which is
consistent with the fact that only few of the drugs might be associated with
amnesia. At an FDR level of $\alpha=0.05$, the wFDR procedure finds $39$ drugs
to be associated with amnesia, the adaptive BH procedure $36$, and the BH
procedure $36$. Since the wFDR procedure employs weights and re-ranks
p-values, it may claim a different set of drugs to be associated with amnesia
than that found by the BH procedure. In fact, $37$ of the drugs found by the
wFDR procedure are different than those found by the BH procedure.

Drugs found by the wFDR procedure but not by the BH procedure to be associated
with amnesia include ``Sirolimus", ``Vitamin B substances", ``Xipamide" and
``Raloxifene". Sirolimus, also known as ``Rapamycin", is able to induce
amnesia in rats, and the FDA has approved its use on humans for post-traumatic
stress disorder treatment; see, e.g., \cite{glover2010differing} and
\cite{Tischmeyer:2003}. However, none of the methods in \cite{Heller:2012}
identified Sirolimus to be associated with amnesia.
Interestingly, Bupropion is identified by the BH procedure and the two discrete procedures of \cite{Heller:2012} but not by the wFDR procedure
to be associated with amnesia. Note that \cite{chavant2011} reported some evidence on the
association between Bupropion and memory disorders.

\subsection{Application to methylation study}

The data set, available from \cite{Lister:2008}, was obtained on a study on
cytosine methylation in two unreplicated lines of \textit{Arabidopsis
thaliana}, wild-type (Col-0) and mutant defective (Met1-3). Corresponding to
each cytosine, the null hypothesis is ``the cytosine is not differentially
methylated between the two lines". The aim is to identity differentially
methylated cytosines. It is known that the number of differentially methylated
cytosines is relatively large for \textit{Arabidopsis thaliana}, i.e., the
proportion $\pi_{0}$ of true null hypotheses is relatively small compared to
$1$.

There are $22265$ cytosines in each line, and each cytosine in each line has
only one observation, i.e., a discrete count that indicates the level of
methylation. We choose cytosines whose total counts for both lines are greater
than $5$ and whose count for each line does not exceed $25$, so as to
filter out genes with unreliable low counts and to better utilize for multiple
testing the jumps in the discrete p-value distributions corresponding to genes without
large counts. This yields $3525$ cytosines, i.e., $3525$ null hypotheses to
test simultaneously. We model the counts for each cytosine in the two lines by
two Binomial distributions whose numbers of trials are respectively the total
counts for all cytosines in individual lines, and use FET to test the null
hypothesis for each cytosine. Two-sided p-values of the tests are collected,
and the FDR procedures or estimators are applied to these p-values.

Based on the evidence in \cite{Zhang2006} that methylated cytosines may belong
to pericentromeric heterochromatin, repetitive sequences, and regions
producing small interfering RNAs, the wFDR procedure partitions the $3525$
null hypotheses into $3$ groups. The groupwise proportions of true null
hypotheses are estimated to be $0.7602744$, $0.5744432$ and $0.6996581$, which
is consistent with the evidence on the three types of genomic regions for
cytosine methylation. At an FDR level of $\alpha=0.05$, the wFDR procedure
finds $449$ differentially methylated cytosines, the adaptive BH procedure
$432$, and the BH procedure $326$.

\section{Discussion\label{Sec:Discussion}}

We have proposed a new grouping algorithm and a weighted FDR procedure for
multiple testing in the discrete paradigm where p-value distributions are
discrete and heterogeneous and null p-values are super-uniform. Under this setting, we have theoretically justified that the wFDR
procedure is non-asymptotically conservative under independence, regardless of how the null hypotheses are partitioned into groups or the feasible value of its tuning parameter.
For multiple testing based on p-values of the Binomial test (BT) and
Fisher's exact test (FET), we have empirically shown that the wFDR procedure is considerably more powerful than six other FDR
procedures under independence, and provided numerical evidence on its non-asymptotic conservativeness and competitive power performance (relative to these other procedures) under positive dependence.
With regard to the number of groups to partition the hypotheses into,
we suggest using domain knowledge on the data for this purpose.
If such knowledge is not available, we recommend
choosing a number that balances the degree of heterogeneity of p-value
distributions within each group and the minimal group size.

We did not study the asymptotic conservativeness of the wFDR procedure under the ``weak dependence" assumption proposed by \cite{Storey:2004}. Once the wFDR procedure partitions hypotheses into groups, this assumption equivalently requires that the empirical distributions of the null p-values and alternative p-values for each group respectively converge
almost surely to two continuous functions, and that the proportion of true null hypotheses for each group converges. In the discrete paradigm of multiple testing, we prefer not to employ such an assumption since it is usually invalid. However, assuming
the assumption is sensible, the asymptotic conservativeness of the wFDR procedure follows directly from that of the data-adaptive one-way GBH procedure of \cite{Nandi:2018}.

The method we have used to construct data-adaptive weights uses a tuning parameter. This
raises the interesting task of finding the values of the tuning parameter and the number of groups that maximize the power of the wFDR procedure subject to a nominal FDR upper bound.
Further, in our simulation studies, we have observed considerable power improvement of the wFDR procedure over the procedures of \cite{Habiger:2015}, \cite{Dohler:2017} and \cite{Chen:2018d}. However, it is challenging to derive simple, analytic conditions for which the wFDR procedure has no less rejections than these three procedures since the former uses discrete p-values and the latter three use randomized p-values, a complicated step-up critical sequence, or mid p-values, respectively. Despite this, we can still check which among them is more powerful on a case-by-case basis.
Finally, computing the supremum norms of the differences between discrete, heterogeneous p-value CDFs in order to implement the new grouping algorithm may be take some time when the number of hypotheses to test and the support of each discrete CDF are very large. So, it is important to identify other types of discrete and heterogeneous p-values distributions (different from those of BTs and FETs) for which these supremum norms can be fast computed (either approximately or equivalently).
We leave these for future research.

\section*{Acknowledgements}

This research was funded by a National Science Foundation Plant Genome
Research Program grant (No. IOS-1025976) to R.W. Doerge, and part of it was
carried out when X. Chen was at the Department of Statistics, Purdue
University. The authors are grateful to John D. Storey for helpful comments.

\section*{Conflict of interest}

The authors have declared no conflicts of interest.

\bibliographystyle{apalike}


\appendix

\section{Proof of \autoref{ThmWFDR}} \label{secProofs}

The following notations and conventions will be used in the proofs of this section and \autoref{secPluginWfdr}: the indicator function $\mathbf{1}_{A}$ of a set $A$ will be
written as $\mathbf{1}A$ if $A$ is described by a proposition.
The $m$ null hypotheses $\left\{  H_{i}\right\}  _{i=1}^{m}$ are partitioned into $l_{\ast}$
disjoint groups $H_{j_{k}},k\in G_{j}$, so are the $m$ p-values $\left\{
p_{i}\right\}  _{i=1}^{m}$ into $p_{j_{k}},k\in G_{j}$, and so are the $m$
weighted p-values $\left\{  \tilde{p}_{i}\right\}  _{i=1}^{m}$ into $\tilde
{p}_{j_{k}},k\in G_{j}$, where $\left\{  G_{j}\right\}  _{j=1}^{l_{\ast}}$
partition the set $\left\{  1,\ldots,m\right\}  $. Recall $I_{0}$ as the index
set of true null hypotheses among the $m$ null hypotheses. For each $G_{j}$,
let $S_{j0}=G_{j}\cap I_{0}$, i.e., $S_{j0}$ is the index set of true null
hypotheses for group $G_{j}$, and $n_{j0}$ be the cardinality of $S_{j0}$. Let $\mathbf{p}=\left(  p_{1},\ldots,p_{m}\right)  $ be the
vector of the $m$ p-values. For each $1\leq i\leq m$, let $\mathbf{p}_{-i}$ be the vector obtained by
removing $p_{i}$ from $\mathbf{p}$, and $\mathbf{p}_{0,i}$ the vector
obtained by setting $p_{i}=0$ in $\mathbf{p}$.

Now we present the arguments.
The proof of the first claim is very straightforward by noticing that the adaptive
procedure of \cite{Chen:2016} is equivalent to
applying the BH procedure to weighted p-values with the weight $\hat{\pi}%
_{0}^{\left(  1\right)  }$.
So, larger weights lead to less rejections, and the claim is valid.

Next we deal with the claim on conservativeness of the wFDR procedure. This part of the proof can be simplified if we use some arguments from the proof of Theorem 2 of \cite{Nandi:2018}, which depends on Theorem 2.1 and Theorem 3.3 in \cite{Sarkar:2008b}. However, for completeness, we will provide streamlined, self-contained arguments here.
Let $\hat{\alpha}$ be the FDR of the wFDR\ procedure at nominal FDR\ level
$\alpha\in\left(  0,1\right)  $, i.e., $\alpha$ is the nominal FDR\ level at
which the BH procedure is applied to the weighted p-values $\left\{  \tilde
{p}_{i}\right\}  _{i=1}^{m}$. For each $1\leq k\leq m$ and $1\leq j\leq
l_{\ast}$, define $R^{\left(  -k\right)  }\left(  \lambda\right)  =\sum
_{i\in\left\{  1,\ldots,m\right\}  \setminus\left\{  k\right\}  }%
\mathbf{1}_{\left\{  p_{i}\leq\lambda\right\}  }$, $R_{j}^{\left(  -k\right)
}\left(  \lambda\right)  =\sum_{i\in G_{j}\setminus\left\{  k\right\}
}\mathbf{1}_{\left\{  p_{i}\leq\lambda\right\}  }$ and%
\[
w_{j}^{\left(  -k\right)  }=\frac{\left(  n_{j}-R_{j}^{\left(  -k\right)
}\left(  \lambda\right)  \right)  \left(  R^{\left(  -k\right)  }\left(
\lambda\right)  +l_{\ast}\right)  }{m\left(  1-\lambda\right)  \left(
R_{j}^{\left(  -k\right)  }\left(  \lambda\right)  +1\right)  }.
\]
Recall $w_{j}$ defined by (\ref{eqe2a}) as %
\[
w_{j}=\frac{\left(  n_{j}-R_{j}\left(  \lambda\right)  +1\right)  \left(
R\left(  \lambda\right)  +l_{\ast}-1\right)  }{m\left(  1-\lambda\right)
R_{j}\left(  \lambda\right)  }.
\]
Since $w_j$ is non-decreasing in $p_{j_k}$, setting $p_{j_k}=0$ when computing $w_j$ directly gives $w_{j}^{\left(  -k\right)  }$.
So, $w_{j}\geq w_{j}^{\left(  -k\right)  }$ for each $j$ and $k$.

Let $R$ be
the number of rejections of the wFDR\ procedure, which will be written as
$R\left(  p_{j_{k}},\mathbf{p}_{-j_{k}}\right)  $ for each $j$ and $k$
whenever needed.
When the p-values are independent and the null p-values are
super-uniform,%
\begin{align}
&  \hat{\alpha}=\mathbb{E}\left[  \sum_{j=1}^{l_{\ast}}\sum_{k\in S_{j0}}%
\frac{1}{R\left(  p_{j_{k}},\mathbf{p}_{-j_{k}}\right)  }\times\mathbf{1}%
  \left\{  p_{j_{k}}\leq\frac{\alpha R\left(  p_{j_{k}},\mathbf{p}%
_{-j_{k}}\right)  }{mw_{j}}\right\}    \right]  \label{eqBnd4}\\
&  \leq\mathbb{E}\left(  \sum_{j=1}^{l_{\ast}}\sum_{k\in S_{j0}}%
\mathbb{E}\left[  \left.  \frac{1}{R\left(  p_{j_{k}},\mathbf{p}_{-j_{k}%
}\right)  }\times\mathbf{1}  \left\{  p_{j_{k}}\leq\frac{\alpha R\left(
p_{j_{k}},\mathbf{p}_{-j_{k}}\right)  }{mw_{j}^{\left(  -k\right)  }}\right\}
  \right\vert \mathbf{p}_{-j_{k}}\right]  \right)  \label{eqBnd3}\\
&  \leq\frac{\alpha}{m}\sum_{j=1}^{l_{\ast}}\mathbb{E}\left[  \sum
\nolimits_{k\in S_{j0}}\frac{1}{w_{j}^{\left(  -k\right)  }}\right]
,\label{eqBnd2}%
\end{align}
where from (\ref{eqBnd4}) to (\ref{eqBnd3}) we have used $w_{j}\geq w_{j}^{\left(  -k\right)  }$ and then conditioned on $\mathbf{p}_{-j_{k}}$, and from (\ref{eqBnd3}) to (\ref{eqBnd2}) we have applied Lemma
4.2 of \cite{Sarkar:2008b} for super-uniform null p-values to the conditional expectation in (\ref{eqBnd3}) and used the fact that
$R\left(p_{j_{k}},\mathbf{p}_{-j_{k}}\right)$ conditional on $\mathbf{p}_{-j_{k}}$ (and hence equivalently conditional on $w_{j}^{\left(  -k\right)  }$ for all $1 \le j \le l_{\ast}$ and a fixed $k$) is non-increasing in $p_{j_{k}}$. Therefore, it
suffices to show%
\begin{equation}
\sum_{j=1}^{l_{\ast}}\mathbb{E}\left[  \sum\nolimits_{k\in S_{j0}}\left(
w_{j}^{\left(  -k\right)  }\right)  ^{-1}\right]  \leq m,\label{eqBnd1}%
\end{equation}
which implies $\hat{\alpha}\leq\alpha$, i.e., the wFDR\ procedure is conservative.

Let $V_{j}\left(  \lambda\right)  =\sum_{i\in S_{j0}}\mathbf{1}_{\left\{
p_{i}\leq\lambda\right\}  }$. Then $n_{j0}-V_{j}\left(  \lambda\right)  \leq
n_{j}-R_{j}\left(  \lambda\right)  $. Let $g$ be a non-negative, real-valued,
measurable function of $R_{j}\left(  \lambda\right)  $. Then,%
\begin{align}
\mathbb{E}\left[  g\left(  R_{j}\left(  \lambda\right)  \right)  \right]    &
\geq\mathbb{E}\left[  \frac{\left(  n_{j0}-V_{j}\left(  \lambda\right)
\right)  g\left(  R_{j}\left(  \lambda\right)  \right)  }{\max\left\{
n_{j}-R_{j}\left(  \lambda\right)  ,1\right\}  }\right]  \nonumber\\
& =\sum_{r^{\prime}=0}^{n_{j}}\sum_{k\in S_{j0}}\mathbb{E}\left[
\frac{\mathbf{1}_{\left\{  p_{j_{k}}>\lambda\right\}  }1_{\left\{
R_{j}\left(  \lambda\right)  =r^{\prime}\right\}  }g\left(  r^{\prime}\right)
}{\max\left\{  n_{j}-r^{\prime},1\right\}  }\right]  \nonumber\\
& \geq\sum_{r^{\prime}=0}^{n_{j}-1}\left(  1-\lambda\right)  \sum_{k\in S_{j0}%
}\mathbb{E}\left[  \mathbf{1}  \left\{  p_{\left(  j,r^{\prime}\right)
}^{\left(  -k\right)  }\leq\lambda<p_{\left(  j,r^{\prime}+1\right)
}^{\left(  -k\right)  }\right\}    \frac{g\left(  r^{\prime}\right)
}{\max\left\{  n_{j}-r^{\prime},1\right\}  }\right]  \label{eqBnd5}\\
& =\sum_{r^{\prime}=0}^{n_{j}-1}\left(  1-\lambda\right)  \sum_{k\in S_{j0}%
}\mathbb{E}\left[  \frac{\mathbf{1}  \left\{  R_{j}^{\left(  -k\right)
}\left(  \lambda\right)  =r^{\prime}\right\}    }{\max\left\{
n_{j}-R_j^{\left(  -k\right)  }\left(  \lambda\right)  ,1\right\}  }g\left(
R_j^{\left(  -k\right)  }\left(  \lambda\right)  \right)  \right]  \nonumber\\
& =\sum_{k\in S_{j0}}\mathbb{E}\left[  \frac{\left(  1-\lambda\right)
g\left(  R_j^{\left(  -k\right)  }\left(  \lambda\right)  \right)  }%
{n_{j}-R_j^{\left(  -k\right)  }\left(  \lambda\right)  }\right]  ,\nonumber
\end{align}
where $p_{\left(  j,r^{\prime}\right)  }^{\left(  -k\right)  }$ is the
$r^{\prime}$-th largest p-value among $\left\{  p_{i},i\in G_{j}%
\setminus\left\{  {k}\right\}  \right\}  $ with $k\in S_{j0}$ and
$p_{\left(  j,0\right)  }^{\left(  -k\right)  }=0$ and $p_{\left(  j,n_j \right)  }^{\left(  -k\right)  }=\infty$ are set, and the independence
among p-values and super-uniformity of null p-values have been used to obtain
(\ref{eqBnd5}). In other words, we have justified
\begin{equation}
\mathbb{E}\left[  g\left(  R_{j}\left(  \lambda\right)  \right)  \right]
\geq\sum_{k\in S_{j0}}\mathbb{E}\left[  \frac{\left(  1-\lambda\right)
g\left(  R_j^{\left(  -k\right)  }\left(  \lambda\right)  \right)  }%
{n_{j}-R_j^{\left(  -k\right)  }\left(  \lambda\right)  }\right]
\label{eqBnd6}%
\end{equation}
for any non-negative, real-valued, measurable function $g$ of $R_{j}\left(  \lambda\right)  $. For each $j$, take $g$ to be the mapping%
\[
R_{j}\left(  \lambda\right)  \mapsto\frac{R_{j}\left(  \lambda\right)
+1}{R\left(  \lambda\right)  +l_{\ast}} = \frac{R_{j}\left(  \lambda\right)
+1}{R_j\left(  \lambda\right) + \sum_{i \ne j} R_i\left(  \lambda\right)  +l_{\ast}}.
\]
Then (\ref{eqBnd6}) implies%
\begin{align*}
\sum_{j=1}^{l_{\ast}}\mathbb{E}\left[  \sum\nolimits_{k\in S_{j0}%
}\left(  w_{j}^{\left(  -k\right)  }\right)  ^{-1}\right]    & =\sum
_{j=1}^{l_{\ast}}\mathbb{E}\left[  \sum_{k\in S_{j0}}\frac{m\left(
1-\lambda\right)  \left(  R_{j}^{\left(  -k\right)  }\left(  \lambda\right)
+1\right)  }{\left(  n_{j}-R_{j}^{\left(  -k\right)  }\left(  \lambda\right)
\right)  \left(  R^{\left(  -k\right)  }\left(  \lambda\right)  +l_{\ast
}\right)  }\right]  \\
& \leq m\sum_{j=1}^{l_{\ast}}\mathbb{E}\left[  \frac{R_{j}\left(
\lambda\right)  +1}{R\left(  \lambda\right)  +l_{\ast}}\right]  =m,
\end{align*}
justifying (\ref{eqBnd1}). This completes the proof.

\section{wFDR procedure with plug-in estimator of oracle weights}
\label{secPluginWfdr}

In this section, we provide a non-asymptotic FDR upper bound on the wFDR procedure that employs a plug-in proportion estimator to construct data-adaptive weights.
This upper bound also applies to the adaptive GBH of \cite{Hu:2010}.

After the wFDR\ procedure
partitions the hypotheses into $l_{\ast}$ groups, if it employs as the weights
\[
w_{j}=\frac{\hat{\pi}_{j0}\left(  1-\hat{\pi}_{0}^{\ast}\right)  }{1-\hat{\pi
}_{j0}}\mathbf{1}_{\left\{  \hat{\pi}_{j0}\neq1\right\}  }+\infty
\times\mathbf{1}_{\left\{  \hat{\pi}_{j0}=1\right\}  } \quad \text{for} \quad j\in\left\{1,\ldots,l_{\ast}\right\},
\]
where $\hat{\pi}_{0}^{\ast}=m^{-1}\sum\limits_{j=1}^{l_{\ast}}\hat{\pi}%
_{j0}\left\vert G_{j}\right\vert $ and each $\hat{\pi}_{j0}$ is given by the
estimator $\hat{\pi}_{0}^{G}$ in \cite{Chen:2016} as an estimate of the proportion $\pi_{j0}^{\ast}$ of true null hypotheses among the $j$th group of null hypotheses, then we call this version
of the wFDR\ procedure \textquotedblleft plug-in wFDR
procedure\textquotedblright, which is also an adaptive GBH of \cite{Hu:2010}. Note that the plug-in wFDR procedure makes no rejections when $\hat{\pi}_{0}^{\ast}=1$,
and that $\hat{\pi}_{0}^{G}$ is ``reciprocally conservative" as justified by Theorem 2 of \cite{Chen:2016}.

The assumptions we need are stated below:

\begin{itemize}
\item[A0)] $\min\left\{  p_{i}: i\ge1\right\}  >0$ almost surely and $\pi_{0} \in\left(  0,1\right)  $
uniformly in $m$.

\item[A1)] $\left\{  p_{i}\right\}  _{i=1}^{m}$ are mutually independent, the null p-values are super-uniform, and no null p-value has its CDF as a Dirac mass.

\item[A2)] If $\pi_{0} \in\left(
0,1\right)  $ uniformly in $m$, then $\lim_{m\rightarrow\infty}\Pr\left(
\hat{\pi}_{0}^{\ast} > \check{\pi}_{0}\right)  =0$ for a sequence of constants
$\check{\pi}_{0}$ such that $\check{\pi}_{0} \le\pi_{0}$.
\end{itemize}

The first part of assumption A0)
avoids the undetermined operation $0\times\infty$ when a p-value is $0$ and a
weight is $\infty$. It is a very mild requirement in the discrete paradigm and
holds automatically in the continuous paradigm. The second part of A0)
excludes the case $\pi_{0}=0$ or $1$ for some $m$, for which the plug-in wFDR
procedure makes no false discoveries or makes all false discoveries and is
thus conservative or anti-conservative automatically. It is also assumed in
the continuous paradigm and is not restrictive. Assumption A1) is used to
validate Theorem 2 in \cite{Chen:2016} on the estimator $\hat{\pi}_{0}^{G}$ (in
the proof of \autoref{thmConserve} hereunder). The second part of A1) requires that a
null p-value distribution should not be degenerate and is not restrictive, and
it holds automatically in the continuous paradigm. Assumption A2) is technical as will be seen later.

With these preparations, we have:

\begin{theorem}
\label{thmConserve}Let $\alpha_{m}^{\ast}$ be the FDR of the plug-in wFDR
procedure. Under assumptions A0) and A1), there exists a constant $\tilde{\pi
}_{0}\in\left(  0,1\right)  $ possibly dependent on $m$ such that
\begin{equation}
\alpha_{m}^{\ast}\leq\alpha\dfrac{1-\pi_{0}}{1-\tilde{\pi}_{0}}+\Pr\left(
\hat{\pi}_{0}^{\ast}>\tilde{\pi}_{0}\right)  . \label{bndwFDR}%
\end{equation}
If in addition assumption A2) holds, then $\limsup_{m\rightarrow\infty}%
\alpha_{m}^{\ast}\leq\alpha$.
\end{theorem}

The proof of \autoref{thmConserve} is provided in \autoref{ProofSecA2}.
The non-asymptotic upper bound in \eqref{bndwFDR} gives an integrated view on how
the proportion $\pi_{0}$ of true null hypotheses and the conservativeness of
the estimator $\hat{\pi}_{0}^{\ast}$ of $\pi_{0}$ jointly affect the FDR of
the plug-in wFDR procedure. In particular, if $\Pr\left(\hat{\pi}_{0}^{\ast}\leq\pi
_{0}\right) \to 1$, then $\tilde{\pi}_{0}=\pi_{0}$ can be set in
(\ref{bndwFDR}), leading immediately to $\limsup_{m \to \infty}\alpha_{m}^{\ast}\leq\alpha$, and the
plug-in wFDR procedure is conservative asymptotically.
On the other hand, the assumption A2) requires $\lim_{m \to\infty}\left\vert \mathbb{E}\left[
\hat{\pi}_{0}^{\ast}\right]  - \pi_{0} \right\vert = 0$ in view of the
conservativeness of the estimator $\hat{\pi}_{0}^{G}$, i.e., $\mathbb{E}%
\left[  \hat{\pi}_{0}^{\ast}\right]  \ge\pi_{0}$. This requires $\hat{\pi}%
_{0}^{\ast}$ to be very accurate on average when $m$ is large. Nonetheless, the upper bound
in inequality \eqref{bndwFDR} is not necessarily tight since it is obtained
from the self-consistency property (see Definition 3 in
\citealp{Blanchard:2009}) of the plug-in wFDR procedure. So, $\hat{\pi}%
_{0}^{\ast}$ does not have to satisfy A2) to ensure the asymptotic
conservativeness of the plug-in wFDR procedure.

Let us discuss a bit more about assumption A0) and those used by Theorem 4 of
\cite{Hu:2010} to prove the asymptotic conservativeness of GBH.
Firstly, the condition on $\pi_{0}$ in A0) is implied by Condition (3.1)
assumed by this theorem. Secondly, Theorem 4 in \cite{Hu:2010} employed two
additional assumptions, i.e., Conditions (3.2) and (3.3) there, that require,
for each group, the empirical distribution of the null p-values converges
almost surely to the identity function on $[0,1]$ and that the empirical
distribution of the alternative p-values converges almost surely to a
continuous function. However, we prefer not to employ such or similar
assumptions in the discrete paradigm.

\subsection{Proof of \autoref{thmConserve}}

\label{ProofSecA2}

Recall $\alpha_{m}^{\ast}$ as the FDR of the plug-in wFDR procedure at nominal
FDR level $\alpha\in\left(  0,1\right)  $. The rest of the proof is divided
into three steps: (i) filter out irrelevant cases from the analysis; (ii)
obtain a non-asymptotic upper bound on $\alpha_{m}^{\ast}$; (iii) show that
the procedure is asymptotically conservative.

\textbf{Step 1.} Let $R=R\left(  \mathbf{p}\right)  $ be the number
of rejections made by the plug-in wFDR procedure. Then, we can assume
$R\left(  \mathbf{p}\right)  \geq1$. Recall $\pi_{j0}=\left\vert
S_{j0}\right\vert \left\vert G_{j}\right\vert ^{-1}$. If $\pi_{j0}=0$, then
the null hypotheses in group $G_{j}$ are false and do not contribute to any
false discovery, and we can exclude the p-values in group $G_{j}$ from the
analysis. So, we can assume $\pi_{j0} >0$ for each $j$. By assumption A0), no
p-value takes value $0$ and $\pi_{0} \in\left(  0,1\right)  $ uniformly in
$m$. So, a weighted p-value is $0$ if and only if its associated weight is
$0$. Note that by its definition the plug-in wFDR procedure makes no
rejections when $\hat{\pi}_{0}^{\ast}=1$. So, when computing $\alpha_{m}%
^{\ast}$, we can exclude the event $\left\{  \hat{\pi}_{0}^{\ast} = 1\right\}
$. Thus, we can assume $\hat{\pi}_{0}^{\ast}<1$, for which $\hat{\pi}_{j0} <1$
for some $j$ and there exists some constant $\tilde{\pi}_{0} <1$ such that
$\hat{\pi}_{0}^{\ast} \le\tilde{\pi}_{0}$. By its definition, $\hat{\pi}%
_{0}^{\ast}>0$ almost surely. Thus, $\tilde{\pi}_{0} \in\left(  0,1\right)  $.

\textbf{Step 2.} Before we proceed further, we need to set up some notations.
Each $\hat{\pi}_{j0}$,
$j=1,...,l_{\ast}$ and $\hat{\pi}_{0}^{\ast}$ will be written as a function of
the p-values when needed. In particular, for each $i$ and $j$, $\hat{\pi}%
_{j0}\left(  \mathbf{p}_{0,i}\right)  $ denotes the estimator $\hat{\pi}%
_{0}^{G}$ of proportion in \cite{Chen:2016} applied to $\mathbf{p}_{0,i}$. 
Let $V\left(  \mathbf{p}%
\right)  $ be the number of false discoveries of the plug-in wFDR procedure.

Let $\mathcal{B}_{m}$ be the
event $\left\{  \hat{\pi}_{0}^{\ast}\leq\tilde{\pi}_{0}\right\}  $, and
$\mathcal{A}_{m}$ the complement of $\mathcal{B}_{m}$. Then%
\begin{equation}
\alpha_{m}^{\ast}\leq\mathbb{E}\left[  \frac{V\left(  \mathbf{p}\right)
}{R\left(  \mathbf{p}\right)  }\mathbf{1}_{\mathcal{B}_{m}}\right]
+\Pr\left(  \mathcal{A}_{m}\right)  , \label{eqx7a}%
\end{equation}
and it suffices to focus on $\gamma_{m}^{\ast}=\mathbb{E}\left[
\frac{V\left(  \mathbf{p}\right)  }{R\left(  \mathbf{p}\right)  }%
\mathbf{1}_{\mathcal{B}_{m}}\right]  $. Set $\tilde{w}_{j}%
=\hat{\pi}_{j0} (1-\hat{\pi}_{j0})^{-1}$ and $\alpha^{\ast}={\left(  1-\hat
{\pi}_{0}^{\ast}\right)  ^{-1}}{\alpha}$. For each $j\in\left\{
1,...,l_{\ast}\right\}  $ and $k\in S_{j0}$, define%
\begin{equation}
\theta_{j_{k}}=\mathbb{E}\left[  \left.  \frac{\mathbf{1}_{\mathcal{B}_{m}}%
}{R\left(  p_{j_{k}},\mathbf{p}_{-j_{k}}\right)  }\mathbf{1}\left\{  p_{j_{k}%
}\leq\frac{R\left(  p_{j_{k}},\mathbf{p}_{-j_{k}}\right)  \alpha^{\ast}}%
{\tilde{w}_{j}m}\right\}  \right\vert \mathbf{p}_{-j_{k}}\right]  \label{eqx3}%
\end{equation}
and%
\[
q_{j_{k}}=\mathbb{E}\left[  \frac{\mathbf{1}_{\mathcal{B}_{m}}}{R}%
\mathbf{1}\left\{  p_{j_{k}}\leq\frac{R\alpha^{\ast}}{\tilde{w}_{j}m}\right\}
\right]  .\label{eqx11}%
\]
Then%
\begin{equation}
q_{j_{k}}=\mathbb{E}\left[  \theta_{j_{k}}\right]  \text{ \ and \ }\gamma
_{m}^{\ast}\leq\sum_{j=1}^{l_{\ast}}\sum_{k\in S_{j0}}q_{j_{k}},
\label{eqx11a}%
\end{equation}
and it suffices to bound each $q_{j_{k}}$ (or $\theta_{j_{k}}$) in order to
bound $\gamma_{m}^{\ast}$.

Pick a $j$ between $1$ and $l_{\ast}$ and $k\in S_{j0}$. Set $\hat{\pi}_{j0}=\hat{\pi}%
_{j0}\left(  \mathbf{p}\right)  $ and $e_{ji}=(1-\hat{\pi}_{j0}\left(
\mathbf{p}_{0,i}\right)  )\hat{\pi}_{j0}^{-1}\left(  \mathbf{p}_{0,i}\right)$.
Then $\hat{\pi}_{j0}\left(  \mathbf{p}_{0,i}\right)  \leq\hat{\pi}_{j0}\left(
\mathbf{p}\right)  $ for any $i$ implies $\tilde{w}_{j}^{-1}\leq e_{ji}$ for any
$i$. For each $i$ and $j$, set%
\begin{equation}
c_{i}\left(  \alpha\right)  =\frac{e_{ji}}{1-\tilde{\pi}_{0}}\frac{\alpha}%
{m}\text{ \ and }Q_{i}=\left\{  p_{i}\leq\frac{R\left(  p_{i},\mathbf{p}%
_{-i}\right)  \alpha^{\ast}}{\tilde{w}_{j}m}\right\}  \label{eqx12}%
\end{equation}
and
\begin{equation}
\label{eqSuperSet}L_{i} = \left\{  p_{i}\leq R\left(  p_{i},\mathbf{p}%
_{-i}\right)  c_{i}\left(  \alpha\right)  \right\}  .
\end{equation}
Then, $Q_{j_{k}} \cap\mathcal{B}_{m} \subseteq L_{j_{k}}$, and almost surely%
\begin{equation}
\label{eqx13}\left(  \left.  \mathbf{1}_{Q_{j_{k}} \cap\mathcal{B}_{m}%
}\right\vert \mathbf{p}_{-j_{k}}\right)  \leq\left(  \left.  \mathbf{1}%
_{L_{j_{k}}} \right\vert \mathbf{p}_{-j_{k}}\right)  ,
\end{equation}
where we have put parentheses around the conditioned random variables to avoid
notational confusion. Inequality (\ref{eqx13}) together with (\ref{eqx3}) and
\eqref{eqSuperSet} implies%
\begin{equation}
\theta_{j_{k}}\leq\mathbb{E}\left[  \left.  \frac{\mathbf{1}_{L_{j_{k}}}
}{R\left(  p_{j_{k}},\mathbf{p}_{-j_{k}}\right)  }\right\vert \mathbf{p}%
_{-j_{k}}\right]  . \label{eqx12a}%
\end{equation}
As with the proof of \autoref{ThmWFDR}, applying Lemma 4.2 of \cite{Sarkar:2008b} for super-uniform null p-values to (\ref{eqx12a}) gives an upper bound for the term on the right
hand side of (\ref{eqx12a}) as $c_{j_{k}}\left(  \alpha\right)  $. So,%
\begin{equation}
\theta_{j_{k}}\leq c_{j_{k}}\left(  \alpha\right)  . \label{eqx3a}%
\end{equation}

However, assumption A1) validates Theorem 2 of \cite{Chen:2016}, i.e.,
$\mathbb{E}\left[    \hat{\pi}_{j0}^{-1}\left(  \mathbf{p}_{0,j_{k}}\right)
\right]  \leq \pi_{j0}^{-1}$. Thus, the definition of
$c_{i}\left(  \alpha\right)  $ in \eqref{eqx12} implies
\begin{align}
\mathbb{E}\left[  c_{j_{k}}\left(  \alpha\right)  \right]   &  =\frac{\alpha
}{m}\mathbb{E}\left[  \frac{1-\hat{\pi}_{j0}\left(  \mathbf{p}_{0,j_{k}%
}\right)  }{\hat{\pi}_{j0}\left(  \mathbf{p}_{0,j_{k}}\right)  }\frac
{1}{1-\tilde{\pi}_{0}}\right]
  \leq\frac{\alpha}{m}\frac{1-\pi_{j0}}{\pi_{j0}}\frac{1}{1-\tilde{\pi}_{0}}.
\label{eqx12b}%
\end{align}
Recall $q_{j_{k}}=\mathbb{E}\left[  \theta_{j_{k}}\right]  $. Combining
(\ref{eqx11a}), (\ref{eqx3a}) and (\ref{eqx12b}) gives%
\begin{equation}
\gamma_{m}^{\ast}\leq\sum_{j=1}^{l_{\ast}}\sum_{k\in S_{j0}}\frac{\alpha}%
{m}\frac{1-\pi_{j0}}{\pi_{j0}}\frac{1}{1-\tilde{\pi}_{0}}=\alpha\frac
{1-\pi_{0}}{1-\tilde{\pi}_{0}}. \label{eqx3b}%
\end{equation}
Combining (\ref{eqx7a}) and (\ref{eqx3b}), we have
\begin{equation}
\label{bndGrand}\alpha_{m}^{\ast}\leq\alpha\frac{1-\pi_{0}}{1-\tilde{\pi}_{0}%
}+\Pr\left(  \mathcal{A}_{m}\right)  ,
\end{equation}
justifying the non-asymptotic upper bound on $\alpha_{m}^{\ast}$.

\textbf{Step 3.} Under the additional assumption A2), we can identify
$\tilde{\pi}_{0}$ in \textbf{Step 2} as the constant $\check{\pi}_{0}$, i.e.,
set $\tilde{\pi}_{0}=\check{\pi}_{0}$ and still maintain the validity of
(\ref{bndGrand}). Due to A2), $\lim_{m\rightarrow\infty} \Pr\left(
\mathcal{A}_{m}\right)  =0$ and $\frac{1-\pi_{0}}{1-\check{\pi}_{0}}%
=\frac{1-\pi_{0}}{1-\tilde{\pi}_{0}}\leq1$. So, (\ref{bndGrand}) implies
$\limsup_{m\rightarrow\infty}\alpha_{m}^{\ast}\leq\alpha$. This completes the
whole proof.

\end{document}